\documentclass[prd,onecolumn]{revtex4}
\usepackage{dcolumn}
\usepackage{multirow}
\usepackage{graphicx}
\usepackage{amssymb}
\usepackage{bm}
\usepackage{hyperref}
\usepackage{epstopdf}
\usepackage{color}
\usepackage{mathrsfs}
\usepackage{amsmath,amssymb,amsthm}
\usepackage{rotating}
\usepackage{sverb, longtable}
\usepackage{subfigure}
\begin{document}
\title{Weak equivalence principle, swampland and $H_0$ tension with fast single radio bursts FRB 180924 and FRB 190523}
\author{Deng Wang}
\email{cstar@sjtu.edu.cn}
\author{Zhaozhou Li}
\author{Jiajun Zhang}
\affiliation{Department of Astronomy, School of Physics and Astronomy, Shanghai Jiao Tong University, Shanghai 200240, China}

\begin{abstract}
Two new fast single radio bursts FRB 180924 and FRB 190523 well localized to massive galaxies have opened a new window to probe and characterize how cosmic baryons are allocated between galaxies, their surroundings and intergalactic medium. We are motivated by testing Einstein's weak equivalence principle with these two cosmic transients which have accurate redshifts. Using photons with different energies emitted by FRB 180924 and FRB 190523, we obtain the bounds $\Delta\gamma<2.16\times10^{-10}$ and $\Delta\gamma<2.55\times10^{-10}$ by
considering the gravitational potential of the Milk Way.
If using the gravitational potential of the Laniakea supercluster instead of the Milk Way one, we obtain the bounds  $\Delta\gamma<1.06\times10^{-14}$ and $\Delta\gamma<1.26\times10^{-14}$.  Moreover, we estimate the effects of the swampland constant $\lambda$ and Hubble constant $H_0$ on the gravitational delays from single FRBs with known redshifts.

\end{abstract}
\maketitle

\section{Introduction}
In radio astronomy, a fast radio burst (FRB) is a transient radio pulse of length ranging from a fraction of one to several milliseconds, caused by some high-energy astrophysical process or engine not yet identified. Since firstly discovered by Lorimer \textit{et al.} in 2007 \cite{1}, FRBs have opened a new window for us to detect the unseen matter in the universe. As bursts of emission are dispersed and scattered by their passage through ionized material containing intergalactic and intercluster medium, they can potentially be used to detect, study and map the medium residing in a diffuse plasma surrounding and in between galaxies and galaxy clusters. With a subsequent series of events including two repeating ones \cite{2,3}, one can know more about the observational properties of FRBs and rule out many alternative astrophysical FRB models (see \cite{4} for a recent review). In the past several years, the millisecond-duration FRBs have been used to constrain the cosmological parameters \cite{5,6} and test the basic physical law, i.e., Einstein's weak equivalence principle (WEP) \cite{7}, which is one of main pillars of general theory of relativity and can be simply stated as any freely falling, uncharged test body follows a geodesic independent of its internal composition and structure. It indicates that
two different species of massless (rest mass is zero)
neutral particles, or two particles belonging to the same specie with
different energies, if emitted simultaneously from the
same cosmic source and traveling through the same gravitational
fields, should arrive at the earth at the same time. By measuring
the small difference of arrival time of two different particles,
one can test the precision of WEP via the well-known Shapiro time delay effect \cite{8}.

After two important repeating events FRB 121102 and FRB 180814 are, respectively, observed by　 Spitler \textit{et al.} \cite{2} and  Amiri \textit{et al.} \cite{3}, Bannister \textit{et al.} \cite{9} recently report a single one FRB 180924 well localized to a massive galaxy with a signal to noise ratio of 21 in one of the high Galactic latitude. This burst and its host have apparently different properties from the previous localized event FRB 121102. Based on the fact that the integrated electron column density along the line of sight closely matches models of the intergalactic medium (IGM), the discovery of FRB 180924 indicates that some FRBs are very clean probes to detect the baryonic matter of the cosmic web. Subsequently, very excitingly, Ravi \textit{et al.} \cite{10} also report another fast single radio burst FRB 190523 localized to a galaxy with a substantial low star-formation rate, which is a thousand times more massive than the host of FRB 121102. 

In this work, our motivation is to test the accuracy of WEP by using these two new burst FRB 180924 and FRB 190523 with high-precision localization. Meanwhile, we notice that the background evolution of the universe is closely related to the accuracy of WEP. As a consequence, in light of recent 4.4$\sigma$ $H_0$ tension ($H_0$ is Hubble constant) between the directly local measurement \cite{11} and the indirectly global derivation under the assumption of $\Lambda$-cold dark matter ($\Lambda$CDM) model \cite{12}, we attempt to test the effect of $H_0$ on the estimated accuracy of WEP. Subsequently, we also evaluate the impact of the recent proposed swampland criterion \cite{13}, which distinguishes whether an self-consistent effective theory or physical phenomena is compatible with the prediction of string theories, on the precision of WEP. Furthermore, a comprehensive analysis of effects of different parameters on the accuracy of WEP are implemented.       

This work is outlined in the following manner.　In the next section, we introduce the analysis method, previous bounds on the WEP and data. In Section III, we give our results and implement a comprehensive analysis of effects of various parameters including $H_0$ and swampland constant $\lambda$ on the accuracy of WEP. In the final section, discussions and conclusions are presented.     

\section{Method, previous bounds and data}
For cosmic transients including the radical FRBs, the total time delay $\Delta t_{\mathrm{obs}}$ observed by the telescope between two different energy bands should at least include contributions from the following five components:
\begin{equation}
\Delta t_{\mathrm{obs}} = \Delta t_{\mathrm{DM}}+\Delta t_{\mathrm{ini}}+\Delta t_{\mathrm{spe}}+\Delta t_{\mathrm{LIV}}+\Delta t_{\mathrm{gra}}, \label{1}
\end{equation}        
where $\Delta t_{\mathrm{DM}}$ denotes the time delay from dispersed medium by the line-of-sight free electron content, $\Delta t_{\mathrm{ini}}$ the intrinsic emission time delay between two photons, $\Delta t_{\mathrm{spe}}$ the potential time delay from special-relativistic effects due to photons with nonzero rest mass, $\Delta t_{\mathrm{LIV}}$ time delay from Lorentz invariance violation, and $\Delta t_{\mathrm{gra}}$ the difference of the so-called Shapiro time delay between two photons with different energies passing by the gravitational field $V(r)$:
\begin{equation}
\Delta t_{\mathrm{gra}}=\frac{\gamma_1-\gamma_2}{c^3}\int_{r_o}^{r_e}V(r)dr, \label{2}
\end{equation}   
where $c$ is speed of light, $r_o$ and $r_e$ are positions of source and observer and $\gamma_1$ and $\gamma_2$ are parametrized post-Newtonian
(PPN) parameters of two photons emitted by a FRB. In general, the WEP stands for all the metric theories of gravity, and indicates that test particles independent of species or energies  should follow identical geodesics and share the same gravitational time delay. 
It makes all the metric theories of gravity give prediction: $\gamma_1=\gamma_2=1$ \cite{add1}, where subscripts denote two different particles. It is worth noting that the key of testing the accuracy of WEP is to distinguish whether $\Delta t_{\mathrm{gra}}=0$. Moving this test into cosmic platform, one can reasonably neglect contributions from $\Delta t_{\mathrm{spe}}$ and $\Delta t_{\mathrm{LIV}}$ in Eq.(\ref{1}) \cite{7}. By making an assumption $\Delta t_{\mathrm{ini}}>0$, one can easily obtain
\begin{equation}
\Delta t_{\mathrm{obs}}-\Delta t_{\mathrm{DM}}>\frac{\Delta\gamma}{c^3}\int_{r_o}^{r_e}V(r)dr, \label{3}
\end{equation}      
where $\Delta\gamma\equiv\gamma_1-\gamma_2$ is to be confronted with data. Roughly speaking, the gravitational field $V(r)$ consists of contributions from three components, i.e., host galaxy $V_{\mathrm{HG}}(r)$, intergalactic medium $V_{\mathrm{IGM}}(r)$ and our local gravitational field $V_{\mathrm{loc}}(r)$. Since a lack of knowledge of $V_{\mathrm{HG}}(r)$ and $V_{\mathrm{IGM}}(r)$, considering the gravitational potential of the Milk Way $V_{\mathrm{MW}}(r)$ and adopting the so-called Keplerian potential in the data analysis, Eq.(\ref{3}) can be simply expressed as    
\begin{equation}
\Delta\gamma < (\Delta t_{\mathrm{obs}}-\Delta t_{\mathrm{DM}})\left[\frac{GM_{\mathrm{MW}}}{c^3}\ln (\frac{D_\mathrm{L}(z)}{b})\right]^{-1}, \label{4}
\end{equation}
where $G$, $b$ and $D_\mathrm{L}(z)$ are, respectively, the gravitational constant, impact parameter of light rays relative to the Milky Way center and luminosity distance of a cosmic transient at a given redshift $z$. It is easy to see that the upper bound of $\Delta\gamma$ can be obtained if the quantities $\Delta t_{\mathrm{obs}}$, $D_{\mathrm{L}}(z)$ and $b$ being related to a transient are given. Note that if considering a large local gravitational field such as Laniakea supercluster $V_L(r)$, then one should replace the above inequality with Eqs.(3-4) in Ref.\cite{a2}.

It is necessary to review several important bounds on the accuracy of WEP during the past three decades. 
In light of similar estimation method, Longo \cite{14} derived the first upper bound $\Delta\gamma<1.6\times10^{-6}$ by using neutrinos (MeV) and photons (eV) emitted from the supernovae SN 1987A in the Large Magellanic Cloud.
By measuring the time delay of a radar signal, Bertotti \textit{et al.} \cite{15} gave $\gamma-1=(2.1\pm2.3)\times10^{-5}$ from the Doppler tracking of the Cassini spacecraft. Lambert and Le Poncin-Lafitte \cite{16} obtained $\gamma-1=(-0.8\pm1.2)\times10^{-4}$ by using the very-long-baseline radio interferometry to measure the deflect of light rays. By choosing a very special gamma ray burst GRB 090510, Gao \textit{et al.} \cite{17} extended the WEP test to photons with GeV energies and predicted $\Delta\gamma<2\times10^{-8}$. Applying the above analysis method in FRB 110220, Wei \textit{et al.} \cite{7} used an inferred redshift $z=0.81$ acquire a relatively conservative limit $\Delta\gamma<2.52\times10^{-8}$. Using FRB 150814 localized to a galaxy at $z=0.492$, Tingay and Kaplan \cite{18} derived an upper bound $\Delta\gamma<1-2\times10^{-9}$. Very interestingly, after considering the effects of the potential fluctuations from large scale structure (LSS), Nusser \cite{19} remarkably improved the accuracy by at least three orders of magnitude, and predicted $\Delta\gamma<1.4\times10^{-13}$ and $\Delta\gamma<2.4\times10^{-12}$ for FRB 150814 at the $2\,\sigma$ and $3\,\sigma$ confidence level (CL), respectively. Using the  measurements of linear polarization from GRB 110721A and GRB 061122, Wei and Wu \cite{a1} gave tight limits $\Delta\gamma<1.3\times10^{-33}$ and $\Delta\gamma<0.8\times10^{-33}$, respectively. Most recently, Xing {\it et al.} \cite{a2} derived out a bound $\Delta\gamma<2.5\times10^{-16}$ in light of FRB 121102 sub-pulses. A complete summary of previous bounds on WEP can also be found in Tab.1 of Ref.\cite{a2}. In the next section, we will present our results from the latest fast single radio bursts FRB 180924 and FRB 190523.  
 
Here we shall describe in advance the observed data used in this analysis.
FRB 180924 was detected with the Australian Square Kilometer Array Pathfinder (ASKAP) in lower-sensitivity searches at a frequency of 1152 MHz \cite{9}. Rapid multi-wavelength follow-up revealed a fading radio source at J2000 coordinates $\mathrm{RA} = 21\mathrm{h}44\mathrm{m}25.255\mathrm{s}$ and $\mathrm{Dec}=-40^\circ54'00''.9$ at $z=0.3214$. The measured time delay between 1.18 GHz and 1.48 GHz is $0.4$ s, and the impact parameter can be easily worked out $b=6.29$ kpc. At $z=0.660$, FRB 190523 was detected with the Deep Synoptic Array ten-antenna prototype (DSA-10) at a dispersion measure (DM) 760.8(6) pc cm$^{-3}$, and localized to J2000 coordinates $\mathrm{RA} =$ 13:48:15.6(2) and $\mathrm{Dec} = \:$+72:28:11(2). The measured time delay of FRB 190523 is about 0.5 s and its corresponding impact parameter $b=6.98$ kpc. Note that, for these two bursts, we take the possibly best mass estimation of the Milk Way $M_{\mathrm{MW}}=1.5\times10^{12} \, \mathrm{M}_{\odot}$ to date, which is recently obtained by Li \textit{et al.} \cite{20}  
\begin{figure}
	\centering
	\includegraphics[width=8cm, height=8cm]{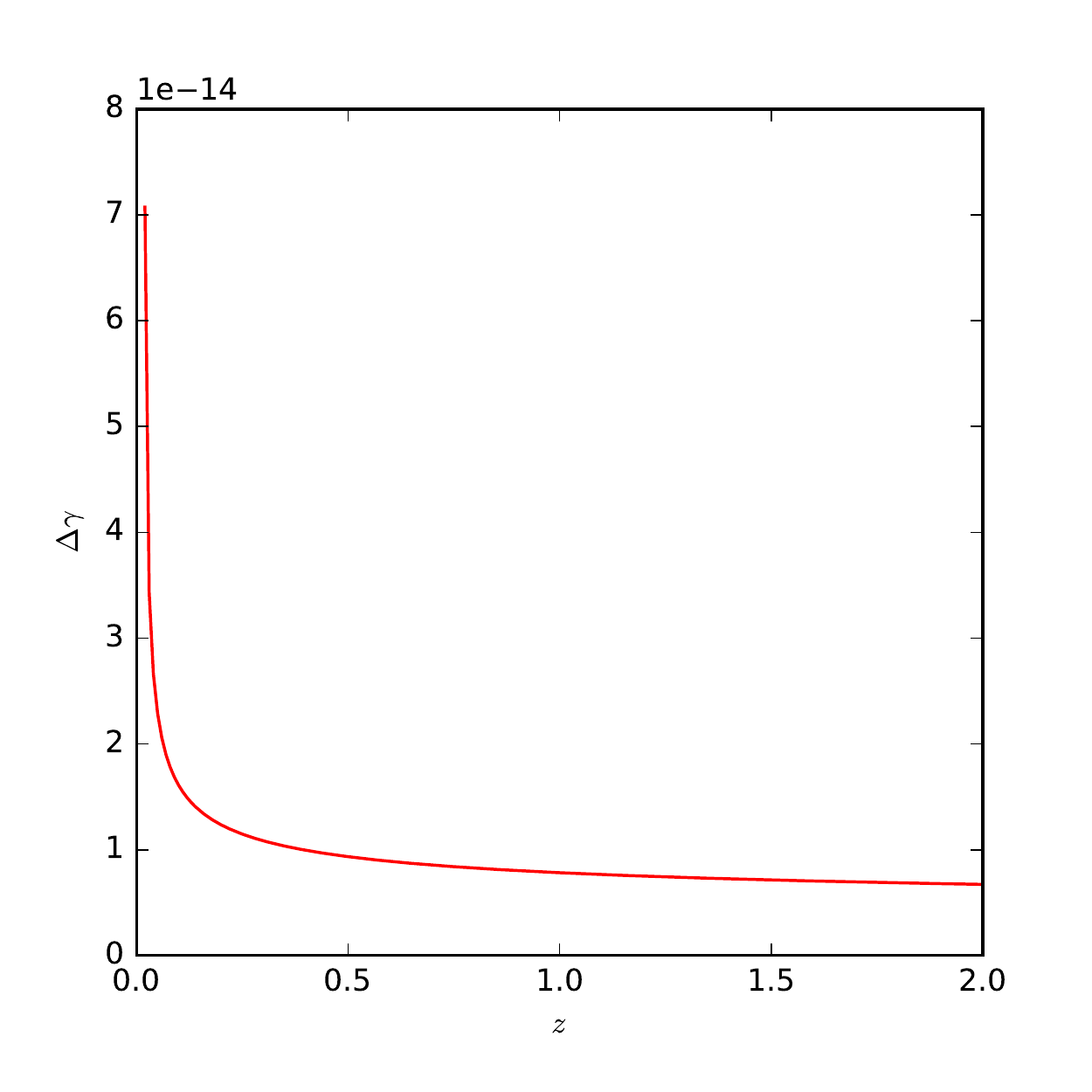}
	\includegraphics[width=8cm, height=8cm]{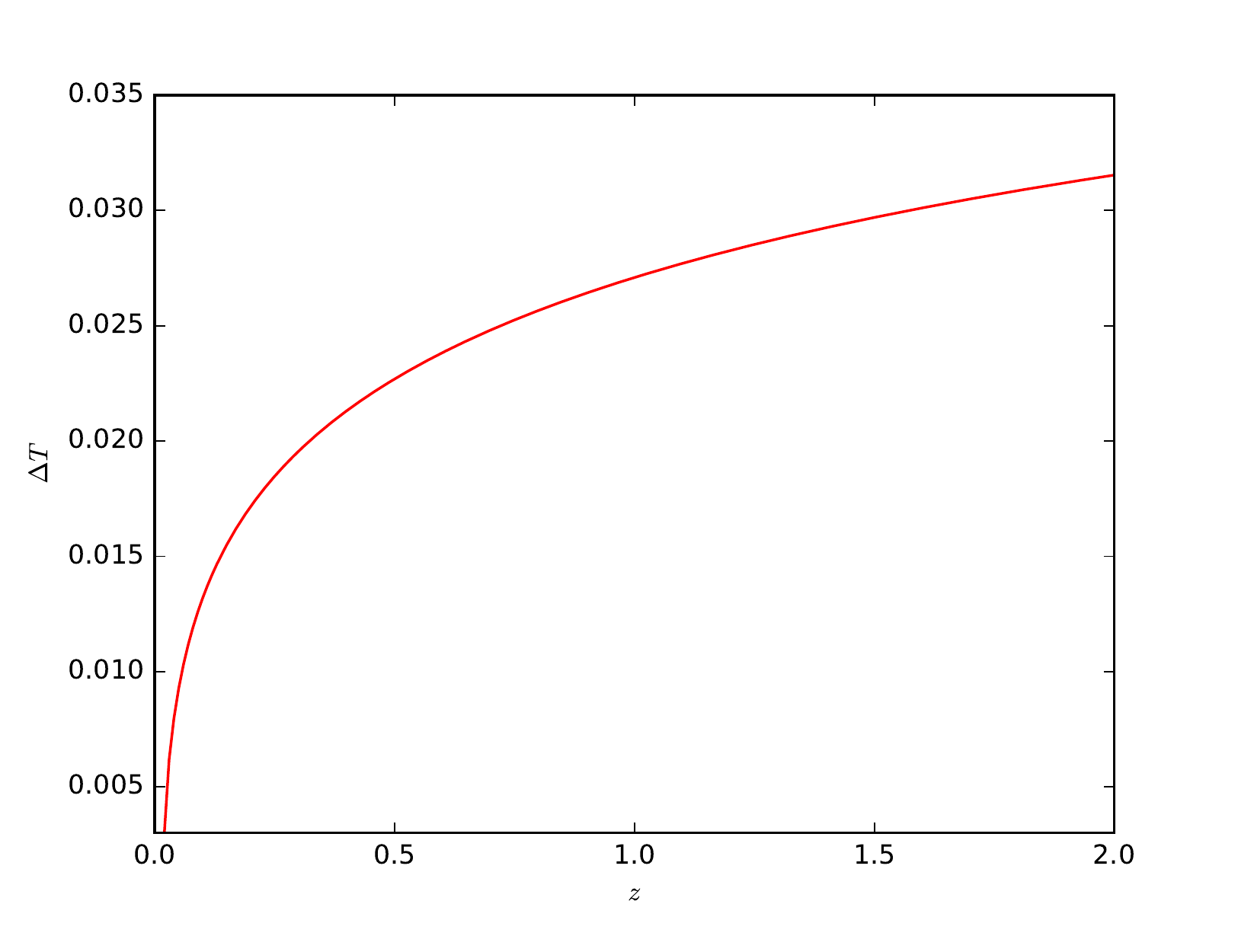}
	\includegraphics[width=8cm, height=8cm]{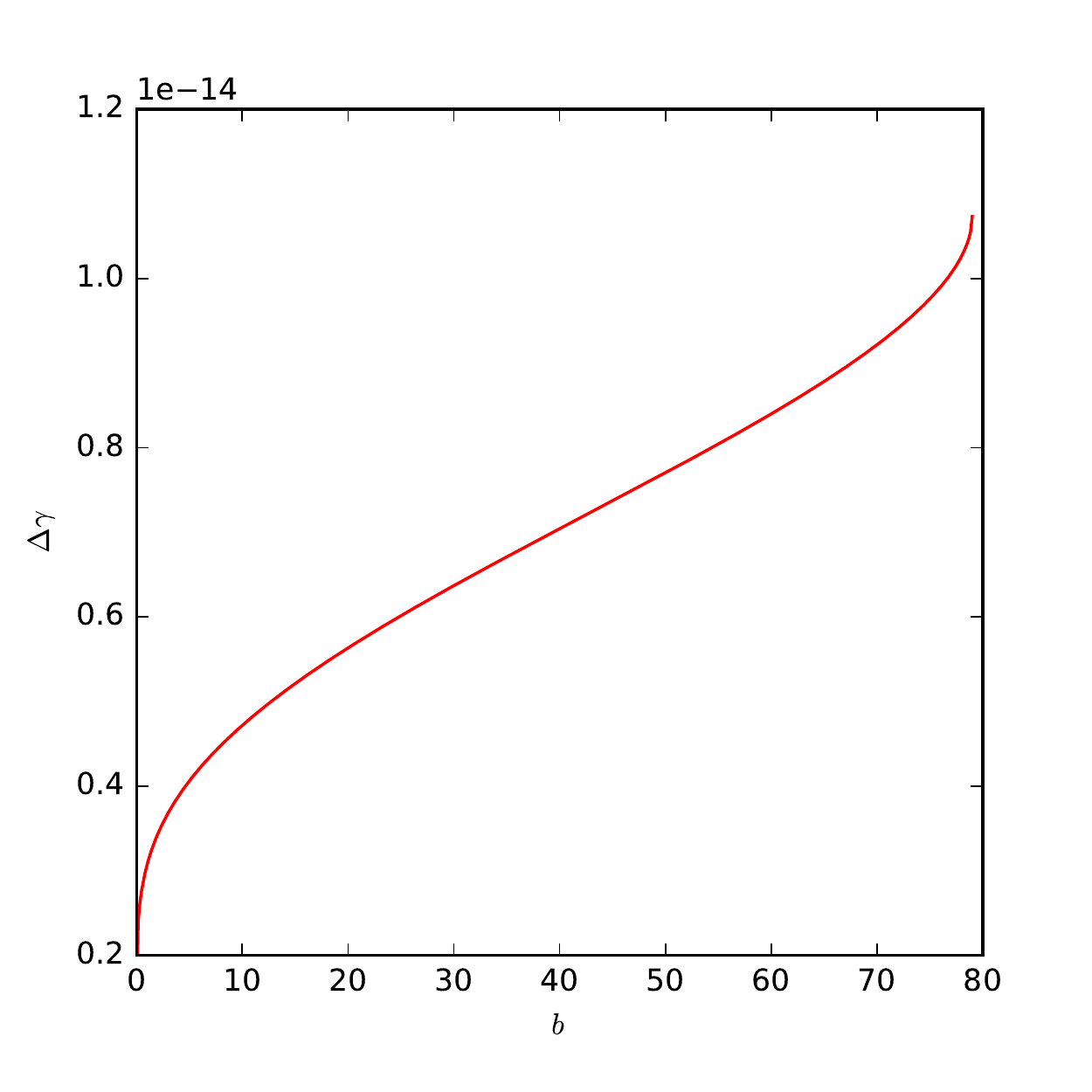}
	\includegraphics[width=8cm, height=8cm]{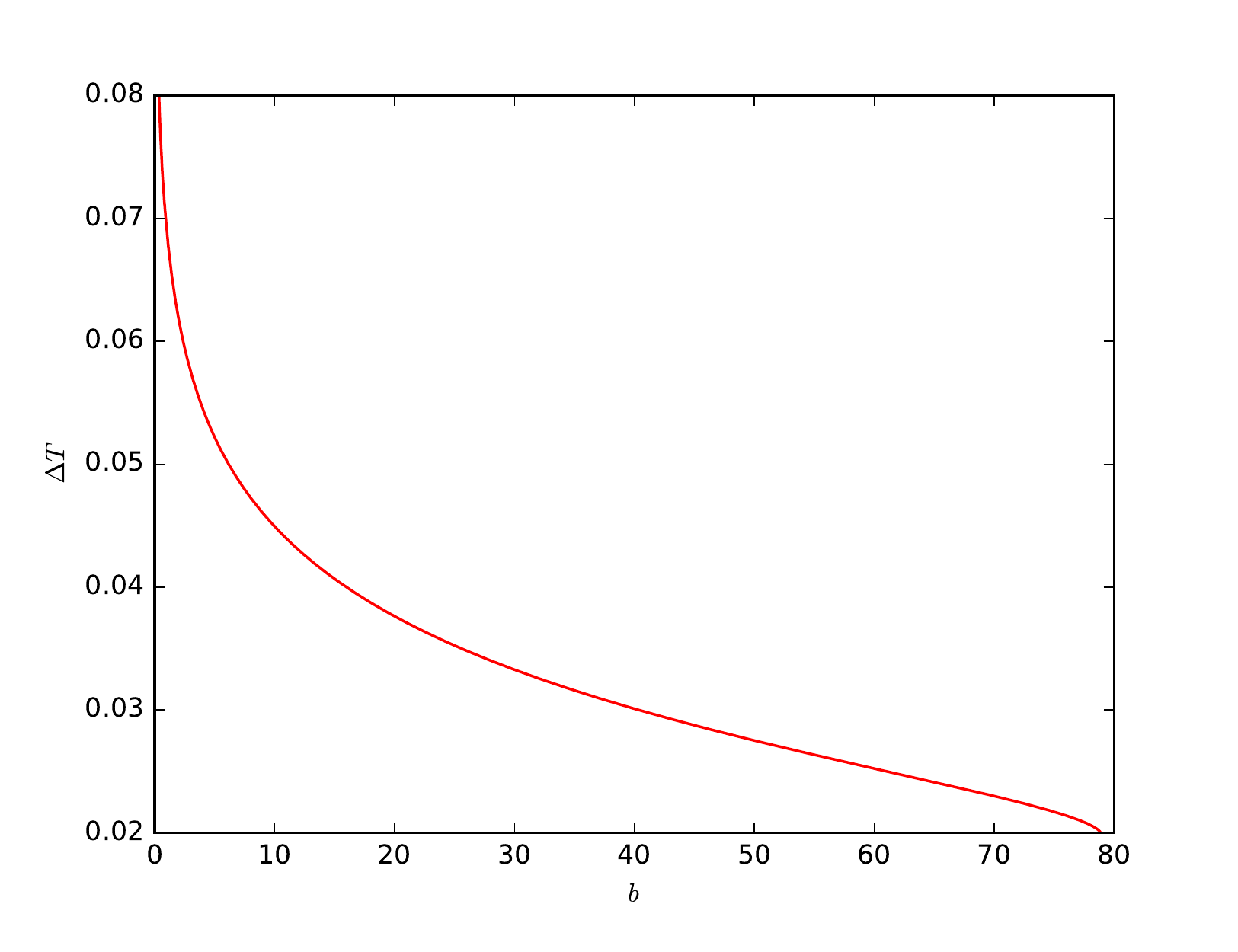}
	\caption{ \emph{Upper left}: the accuracy of WEP $\Delta\gamma$ as a function of redshift $z$ when fixing the impact parameter $b=78.93$ Mpc and $\Delta T=0.02$ s; \emph{Upper right}: the difference $\Delta T$ between gravitational delays $\Delta t_{\mathrm{DM}}$  caused by dispersed medium and observed time delays $\Delta t_{\mathrm{obs}}$ as a function of $z$ when fixing $b=78.93$ Mpc and $\Delta\gamma=1.06\times10^{-14}$; \emph{Lower left}: $\Delta\gamma$ as a function of $b$ when fixing $z=0.3214$ and $\Delta T=0.02$ s; \emph{Lower right}: $\Delta T$ as as a function of $b$  when fixing $z=0.3214$ and $\Delta\gamma=1.06\times10^{-14}$. Here we just consider the case of FRB 180924 and use the $\Lambda$CDM cosmology $\Omega_{m}=0.315$, $\Omega_{\Lambda}=0.685$ and $H_0=67.36$ km s$^{-1}$ Mpc$^{-1}$. }\label{f1}
\end{figure}

\begin{figure}[htbp]
	\centering
	
	\subfigure[\quad swampland]{
		\begin{minipage}[htbp]{0.5\linewidth}
			\centering
			\includegraphics[width=7cm, height=7cm]{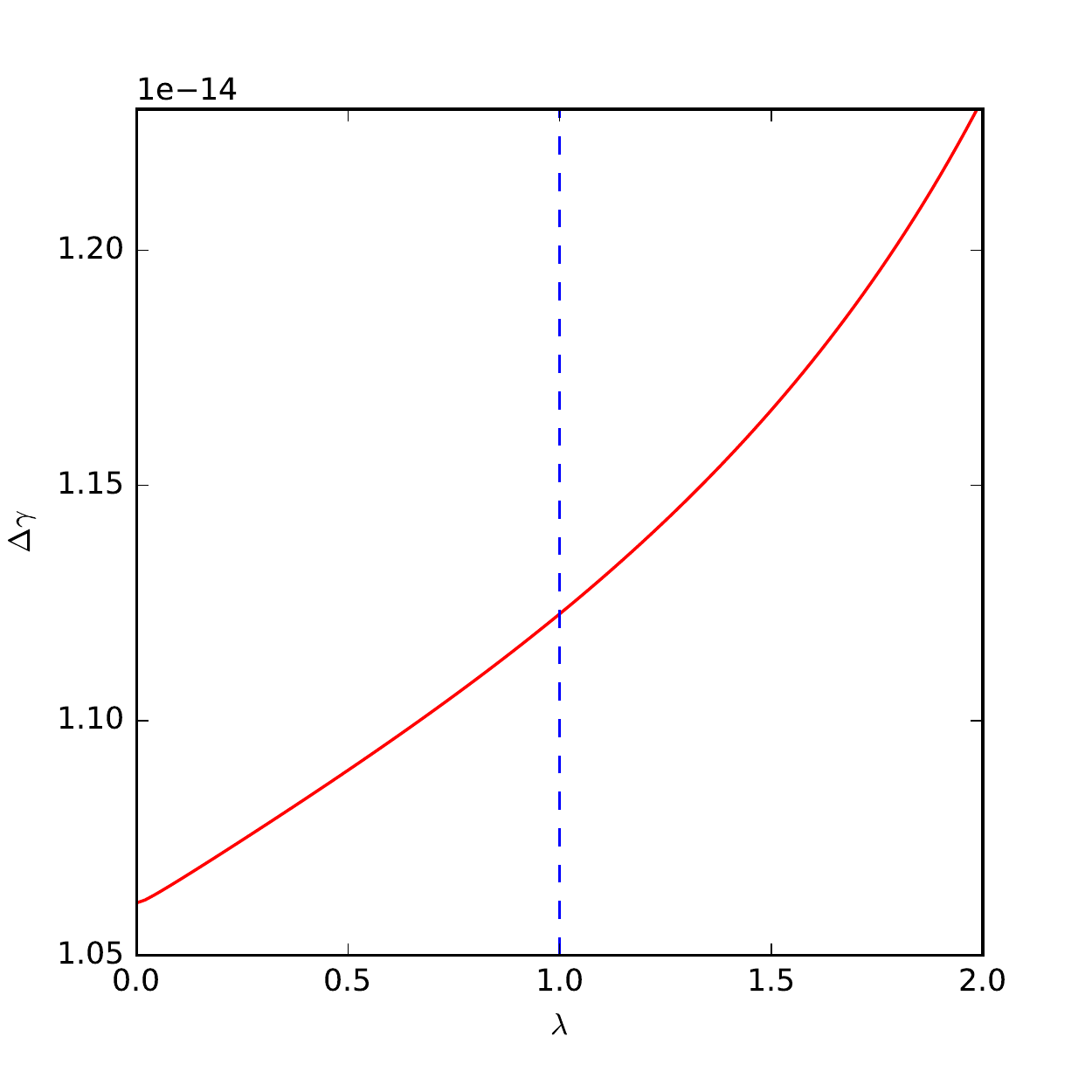}
		\end{minipage}%
	}%
	\subfigure[\quad swampland]{
		\begin{minipage}[htbp]{0.5\linewidth}
			\centering
			\includegraphics[width=7cm, height=7cm]{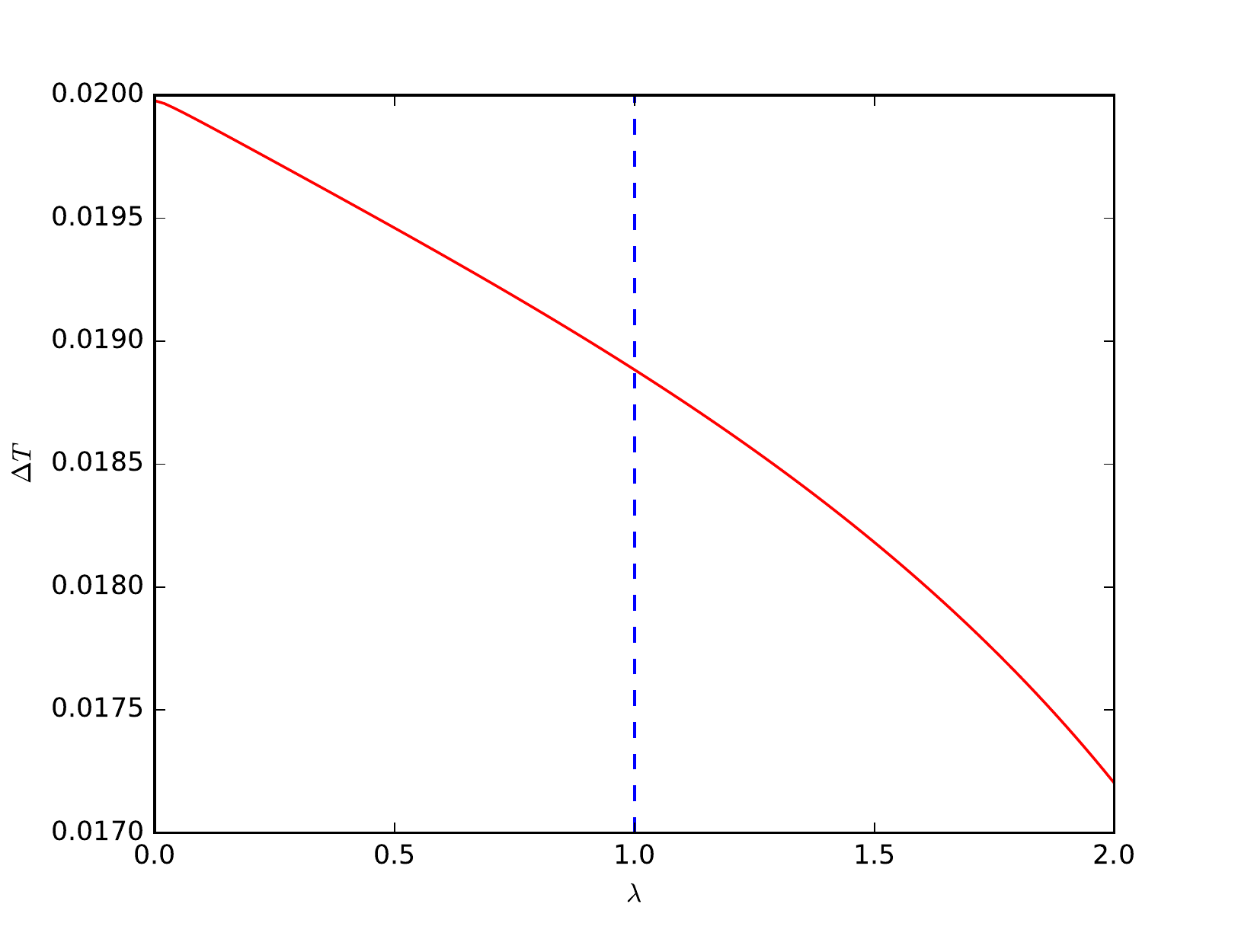}
		\end{minipage}%
	}%
	\\                 
	\subfigure[\quad $H_0$ tension]{
		\begin{minipage}[htbp]{0.5\linewidth}
			\centering
			\includegraphics[width=7cm, height=7cm]{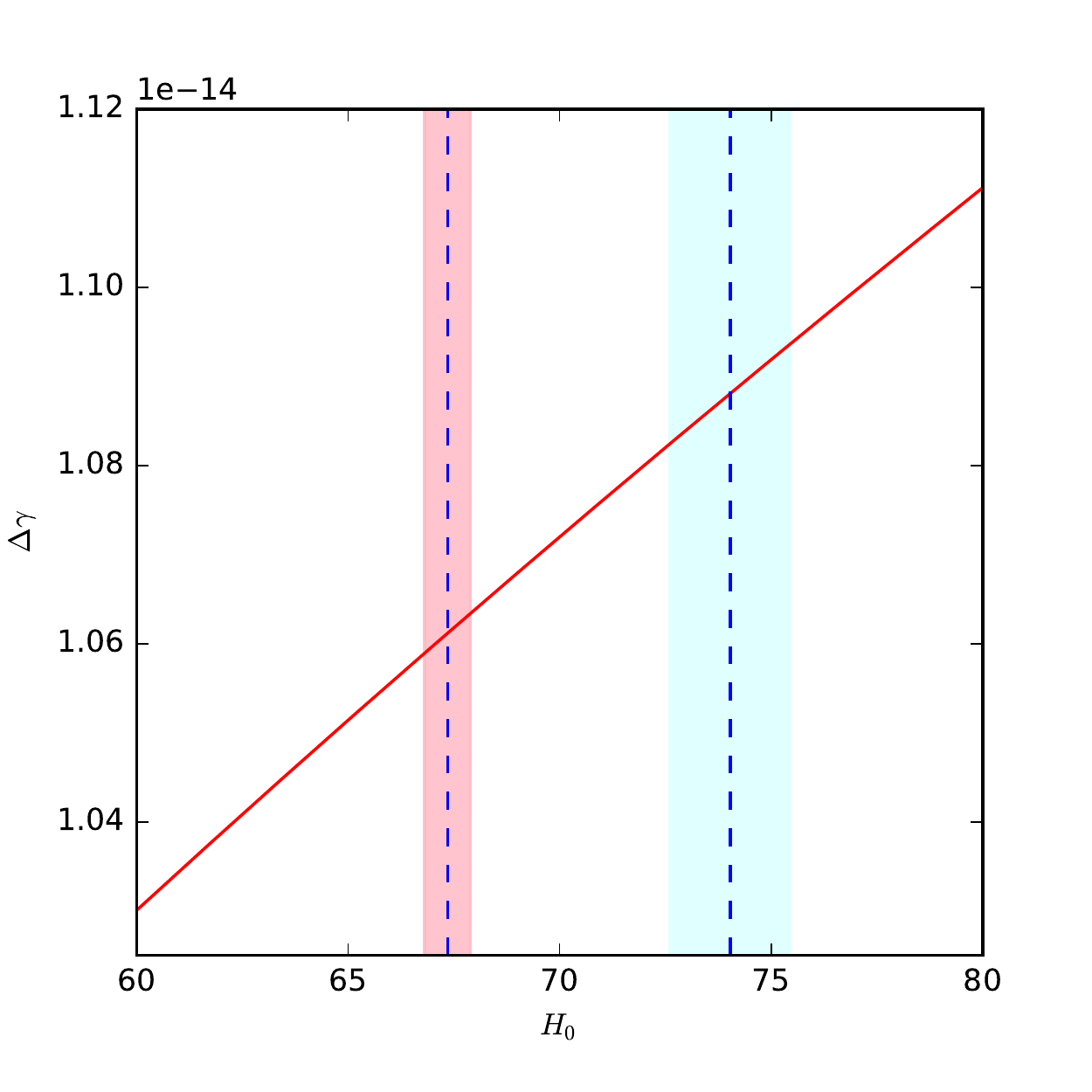}
		\end{minipage}
	}%
	\subfigure[\quad $H_0$ tension]{
		\begin{minipage}[htbp]{0.5\linewidth}
			\centering
			\includegraphics[width=7cm, height=7cm]{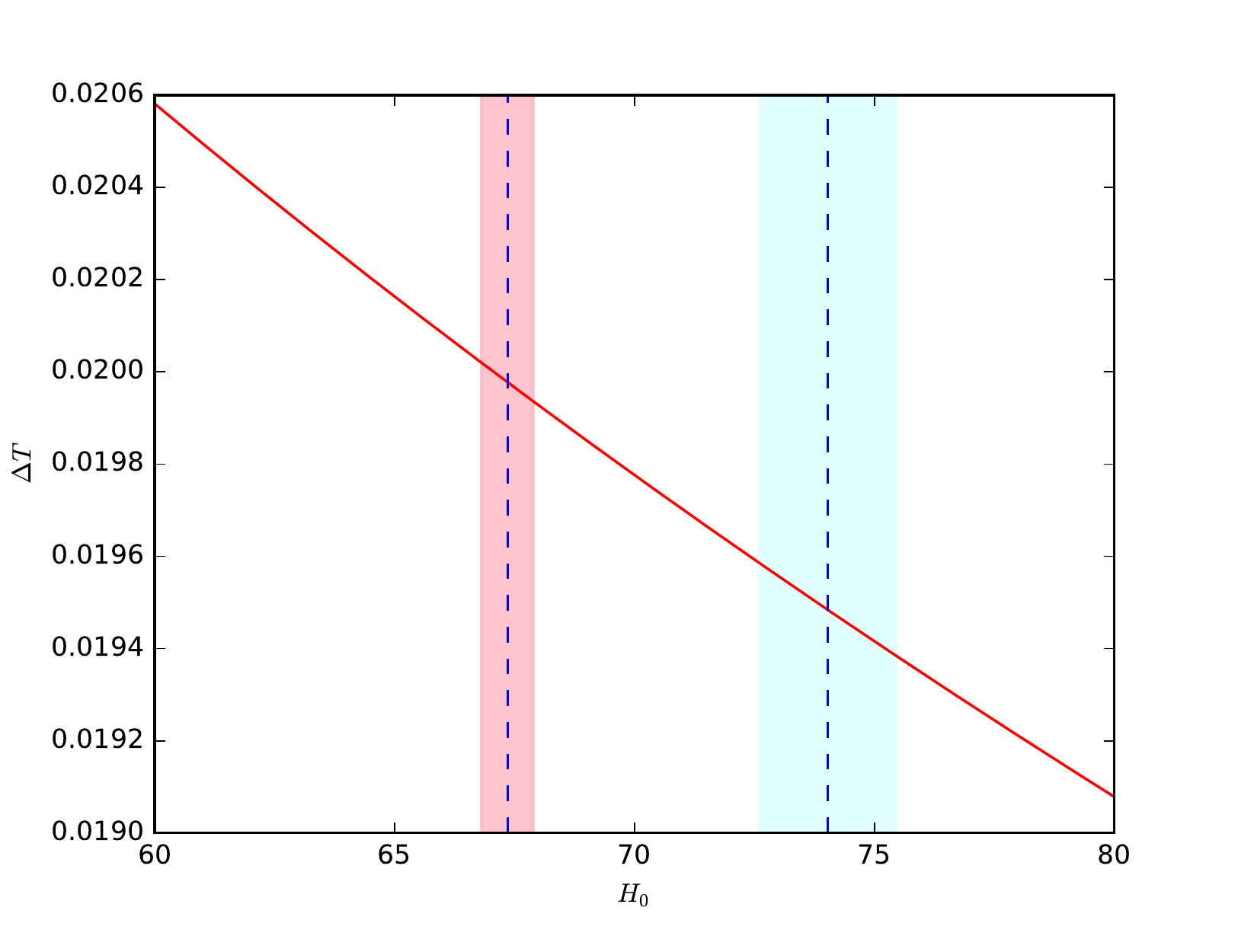}
		\end{minipage}
	}%
	\\                 
	\subfigure[\quad EoS of DE]{
		\begin{minipage}[htbp]{0.5\linewidth}
			\centering
			\includegraphics[width=7cm, height=7cm]{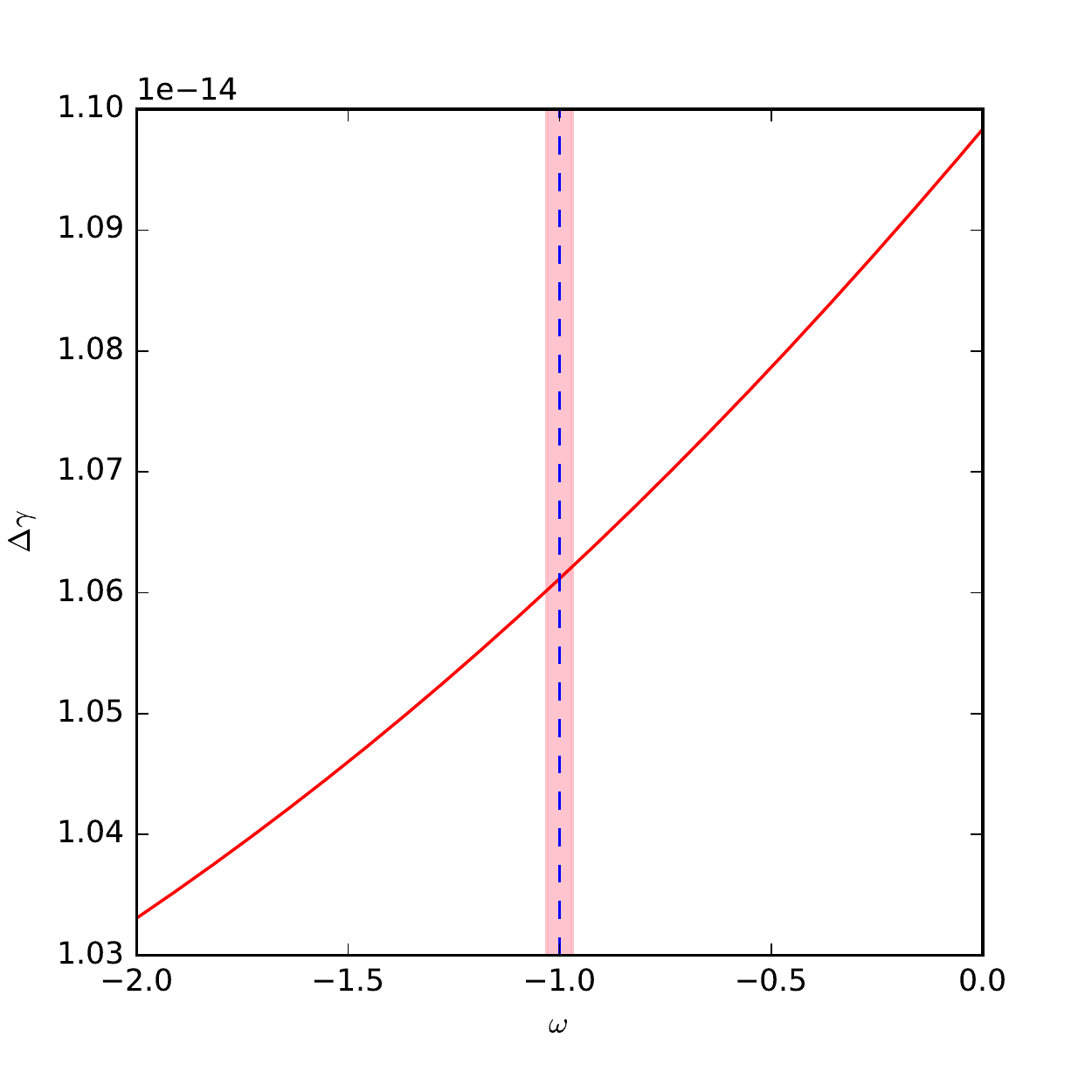}
		\end{minipage}
	}%
	\subfigure[\quad EoS of DE]{
		\begin{minipage}[htbp]{0.5\linewidth}
			\centering
			\includegraphics[width=7cm, height=7cm]{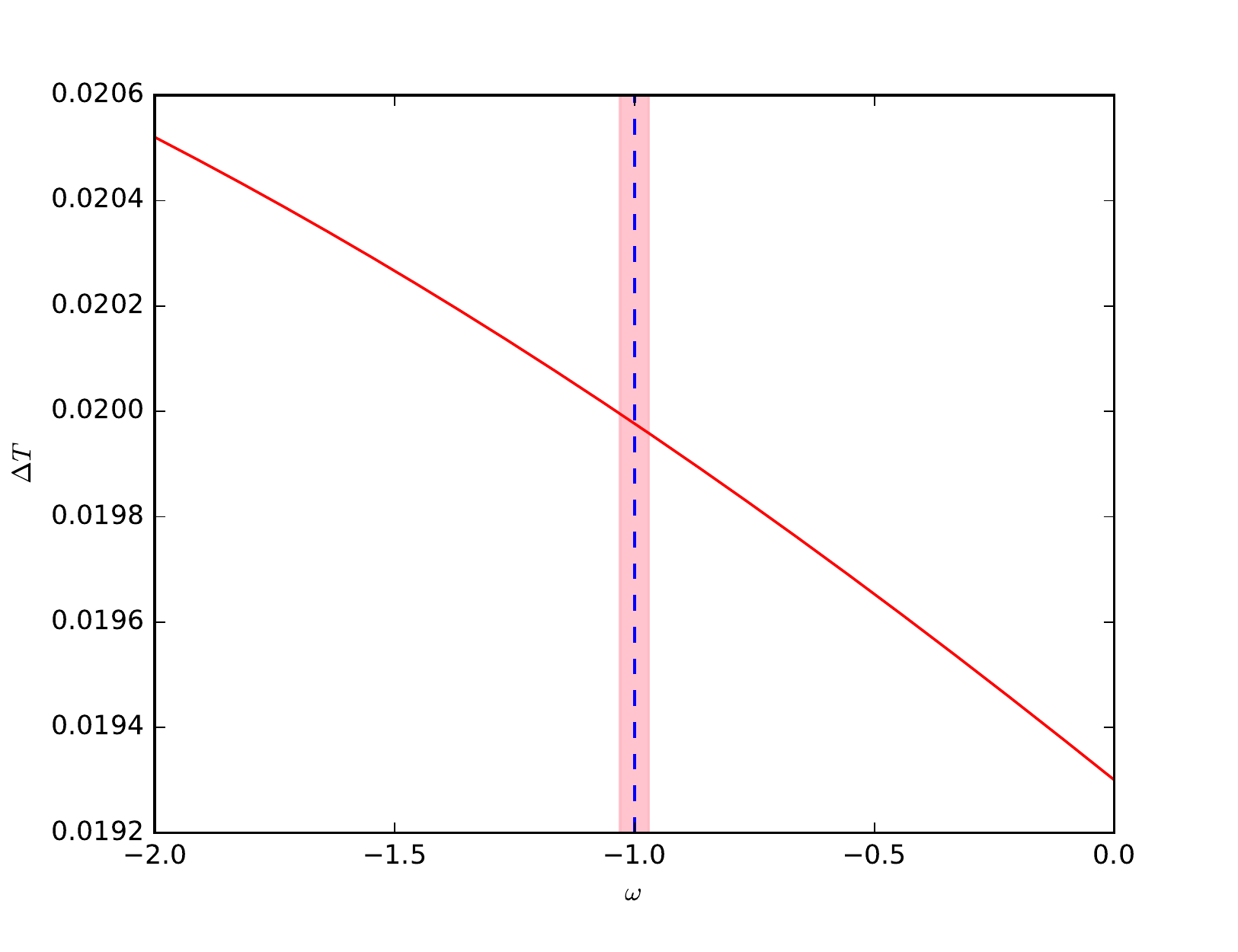}
		\end{minipage}
	}%
	
	\caption{ (a) $\Delta\gamma$ as a function of the swampland constant $\lambda$ when fixing $\Delta T = 0.02$ s;
		(b) $\Delta T$ as a function of $\lambda$ when fixing $\Delta\gamma=1.06\times10^{-14}$;
		(c) $\Delta\gamma$ as a function of the Hubble constant $H_0$ when fixing $\Delta T = 0.02$ s;
		(d) $\Delta T$ as a function of $H_0$ when fixing $\Delta\gamma=1.06\times10^{-14}$;
		(e) $\Delta\gamma$ as a function of a constant equation of state (EoS) of dark energy (DE) $\omega$ when fixing $\Delta T = 0.02$ s;
		(f) $\Delta T$ as a function of $\omega$ $\Delta\gamma=1.06\times10^{-14}$.  
		Here we just consider the case of FRB 180924, use $z=0.3214$ and $b=78.93$ Mpc and adopt the $\Lambda$CDM cosmology $\Omega_{m}=0.315$ and $\Omega_{\Lambda}=0.685$. 
		In two upper panels, the blue dashed lines $\lambda=1$ represents the lower limit predicted by stringy DE models. In two middle panels, two blue dashed lines and corresponding shaded regions from left to right denote $H_0=67.36\pm0.54$ and $H_0=74.03\pm1.42$. In two lower panels, the blue dashed lines and shaded regions mean $\omega=-1\pm0.03$.   
	}\label{f2}
\end{figure}

\section{Results, swampland and $H_0$ tension}
In order to obtain the constraining results, the first step is adopting the reasonable assumptions, i.e., the vanilla $\Lambda$CDM cosmology, matter density ratio $\Omega_{m}=0.315$, DE density ratio $\Omega_{\Lambda}=0.685$, $H_0=67.36$ km s$^{-1}$ Mpc$^{-1}$, and the relatively conservative assumption $\Delta t_{\mathrm{obs}}\gg\Delta t_{\mathrm{DM}}$ as Ref.\cite{7}. The second step is estimating the effect of known redshift for a FRB localized to a massive galaxy on the accuracy of WEP. For FRB 180924, since we find that the errors in the DM of intergalactic medium $\mathrm{DM}_{\mathrm{IGM}}$ and other DM components are also of order $20\%$ like FRB 150814 and even larger (see Fig.S5 in the supplementary materials of Ref.\cite{9}), we use the same analysis method as Ref.\cite{7} to estimate the precision of WEP. If errors in these quantities of $5\%-10\%$ magnitude
masked gravitational time delays, the bound on $\Delta\gamma$ would
be reduced by factors of $10-20$. Therefore, we obtain the limitation $\Delta\gamma<2.16\times10^{-10}$ for a factor of 20, which is, so far, the most stringent constraint from photons with different energies for a non-repeating FRB by only considering the gravitational potential of the Milk Way. Furthermore, if we consider the LSS effect, after some numerical calculations, we have the bounds $\Delta\gamma<2.06\times10^{-13}$ and $\Delta\gamma<3.53\times10^{-12}$ at the $2\,\sigma$ and $3\,\sigma$ CL, respectively. One can find that the choice of gravitational potential affects largely the accuracy of WEP. Following this logical line, we shall also estimate the precision of WEP for non-repeating FRB 180924 by replacing the Milk Way potential with a larger local gravitational field, the Laniakea supercluster, the mass of which is $\sim10^{17}\mathrm{M}_\odot$. Here we use the Great Attractor as gravitational center of Laniakea \cite{21}. After working out the impact parameter $b=78.93$ Mpc according to the formula in Ref.\cite{14} and use Eqs.(3-4) in Ref.\cite{a2}, we obtain the bound $\Delta\gamma<1.06\times10^{-14}$, which is also the strictest bound from photons with different energies for a single FRB by using the Laniakea supercluster to date.

FRB 190523 is also a FRB 150814-like event and they have the almost same observed DM and also relatively large $\mathrm{DM}_{\mathrm{IGM}}$. Similarly, using the above analysis, we obtain the corresponding bound $\Delta\gamma<2.55\times10^{-10}$ for a factor of 20. Similarly, taking the LSS effect into account, we have the limits $\Delta\gamma<2.43\times10^{-13} \, (2\,\sigma)$ and $\Delta\gamma<4.17\times10^{-12} \, (3\,\sigma)$. When using the Laniakea supercluster as the local gravitational engine, we derive $\Delta\gamma<1.26\times10^{-14}$.      

The bound $\Delta\gamma<2.16\times10^{-10}$ for FRB 180924 also implies that $\Delta t_{\mathrm{gra}}<0.02$ s. In light of rapid progress of FRB cosmology, we think that this limit would be smaller and smaller and we shall give a relatively universal upper bound on the accuracy of WEP for a non-repeating FRB.
During the next two decades, if astrophysicists could improve the accuracy of $\mathrm{DM}_{\mathrm{IGM}}$ from $20\%$ to $0.1\%$ (i.e., $0.025\%-0.05\%$ of $\Delta t_{\mathrm{DM}}$ masked the gravitational delays), if they could detect the FRBs out to $z\sim2$,
if the observed time delays can reach the level of $\Delta t_{\mathrm{obs}}\sim0.1$ ms like the repeating burst FRB 121102 ($\Delta t_{\mathrm{obs}}=0.4$ ms \cite{22}), which has a complex time-frequency structure, if the Laniakea supercluster is used, and if millions of observed FRBs from all the directions over the sky make the impact parameter be $b\sim0.01$ kpc, we can obtain the universal bound $\Delta\gamma<8.24\times10^{-22}$ from photons with different energies for all the single FRBs.

To understand this constraining method better, for the first time, we give a comprehensive analysis of the effects of various parameters on the accuracy of WEP $\Delta\gamma$ and the difference $\Delta T= \Delta t_{\mathrm{obs}}-\Delta t_{\mathrm{DM}}$ between gravitational delays caused by dispersed medium and observed time delays. Specifically, we take FRB 180924 in the gravitational field of Laniakea  supercluster as an example to implement this analysis.    

In Fig.\ref{f1}, fixing $\Delta T = 0.02$ s, we find that the WEP accuracy $\Delta\gamma$ decreases very fast when $z<0.2$ but at most one order of magnitude, and then tends to be flat out to high redshifts.  Fixing $\Delta\gamma<1.06\times10^{-14}$, we find that, for FRBs with the same level of WEP precision, their gravitational delays will monotonically increase with $z$. One can also find that, for FRBs with the almost same gravitational delay, their WEP accuracy will increase fast with increasing distances to gravitational center of Laniakea supercluster. 

Considering the quintessence dark energy model with an exponential potential $V(\phi)=V_0e^{\lambda\phi}$ ($V_0$ is the current value of the potential and $\phi$ is quintessence scalar field) as the fiducial cosmology, one can estimate the effect of swampland constant $\lambda$ on the WEP accuracy. Panel (a) in Fig.\ref{f2} shows that, for FRBs with same gravitational delay and close redshifts, their WEP accuracy increases very slowly with increasing $\lambda$, and the lower limit $\lambda=1$ predicted by a self-consistent stringy DE models has a $5.79\%$ effect on the WEP accuracy. 

Panel (b) tells us that for FRBs with same WEP precision, their gravitational delays slowly decreases with $\lambda$, and for a FRB 180924-like event, the validity of swampland criterion can be distinguished as long as its precision of gravitational time delay could be less than $5.48\%$. Panel (c) reflects that the current $H_0$ tension has a $2.53\%$ effect on the WEP precision. Panel (d) indicates that this tension can be distinguishable if the ratio of errors of measured gravitational delays for a single FRB with different energies of photons are smaller than $2.47 \%$. We also exhibit the effects of EoS of DE on WEP accuracy and gravitational time delay in Panels (e) and (f), and find that $\Delta\gamma$ and $\Delta T$ do not have evident dependencies on the EoS of DE based current constraints \cite{12}. In the future, we cannot improve the constraints on the EoS of DE until we are able to measure gravitational delays for non-repeating FRBs within a $0.09\%$ accuracy.        

\section{Discussions and conclusions}
Most recently, two fast single radio bursts FRB 180924 and FRB 190523 well localized to massive galaxies are discovered by two independent groups \cite{9,10}. It is such a big breakthrough since the first FRB is reported by Lorimer \textit{et al.} in 2007 \cite{1}. We are motivated by testing Einstein's WEP with these two new non-repeating bursts which have accurate redshifts.   

Using two photons with different energies emitted by bursts, we obtain the bound on the precision of WEP $\Delta\gamma<2.16\times10^{-10}$ by only considering the gravitational potential of the Milk Way.
If considering the potential fluctuations of LSS, we have the limits $\Delta\gamma<2.06\times10^{-13}$ and $\Delta\gamma<3.53\times10^{-12}$ at the $2\,\sigma$ and $3\,\sigma$ CL, respectively. Furthermore, when using the gravitational potential of the Laniakea supercluster instead of the Milk Way one, we derive out the bound $\Delta\gamma<1.06\times10^{-14}$. Utilizing the same method, for FRB 190523, we obtain the upper bound  $\Delta\gamma<2.55\times10^{-10}$, $\Delta\gamma<2.43\times10^{-13} \, (2\,\sigma)$ and $\Delta\gamma<4.17\times10^{-12} \, (3\,\sigma)$, and $\Delta\gamma<1.26\times10^{-14}$ by considering the Milk Way, LSS potential fluctuations and the Laniakea supercluster as gravitational engines, respectively. Because of rapid progress of FRB observations, we also give an universal bound $\Delta\gamma<8.24\times10^{-22}$ for single FRBs with accurate redshifts.

In theory, it is also very interesting to estimate the effect of the so-called swampland constant $\lambda$ and $H_0$ using these two transients. We estimate the validity of swampland criterion could be distinguished as long as the precision of gravitational time delay from a single FRB with known redshift could be less than $5.48\%$. Meanwhile. $H_0$ tension could be distinguishable (i.e., which value of $H_0$ is preferred) as long as the accuracy of measured gravitational delay for a single FRB with known redshift could be less than $2.47\%$. The constraints on the EoS of DE could be improved unless we have abilities to measure gravitational delays for single FRBs within a $0.09\%$ precision.  This implies that we should measure the dispersion caused by intergalactic medium with a very high precision measurement.  

Moreover, these two new localizations show that FRBs can originate in a diversity of environments, and give a challenge to the previously favored FRB engine model, namely magnetars, which are a type of highly magnetized, rapidly spinning neutron star. Since host galaxies of FRB 180924 and FRB 190523 cannot provide good star-forming environments unlike that of FRB 121102, the prevailing magnetar model needs to be reconsidered. We expect that more data coming soon can help us address this FRB puzzle and tell us how cosmic baryons are allocated between galaxies, their surroundings and intergalactic medium.

\section{Acknowledgements}
Deng Wang thanks Pengjie Zhang, Bin Wang, Hai Yu, Ji Yao and Minji Oh for helpful discussions. 
Deng Wang appreciates useful communications in HOUYI workshop.
Deng Wang is supported by the Super Postdoc Project of Shanghai City.

\end{document}